# Optimized Analog Information Multiplexing Across Custom-Tailored MetaCavity Configurations

*Philipp del Hougne[1]\*, Matthieu Davy[2], Ulrich Kuhl[1]*

[1] Institut de Physique de Nice, CNRS UMR 7010, Université Côte d'Azur, 06108 Nice, France
[2] Univ Rennes, CNRS, Institut d'Electronique et de Télécommunications de Rennes (IETR) - UMR 6164, 35000 Rennes, France
\* philipp.delhougne@inphyni.cnrs.fr

**Combining tunable metasurfaces with chaotic cavities opens new avenues for finely tailored dynamic control of microwaves with programmable coding metacavities (PCMs). There is currently a strong interest in utilizing PCMs to overcome the notorious difficulty of coherent measurements across large apertures at radiofrequencies, with important applications in imaging and sensing for security screening, medical diagnosis and human-computer interaction. Such approaches rely on multiplexing spatially encoded information across a random sequence of PCM coding patterns for single-port single-frequency acquisition. Here, it is shown that a judiciously tailored rather than random coding sequence is necessary to unlock the full potential of PCMs for analog multiplexing. Specifically, the singular value spectrum of the multiplexing channel matrix is tailored to be perfectly flat – as opposed to downward sloping. In-situ experiments show that thereby the number of necessary measurements to achieve a given reconstruction quality is lowered by a factor of 2.5. Computational imaging and other microwave metrology applications are expected to benefit from the resulting reduction in acquisition time, processing burden and latency. The proposed approach and platform also set the stage for future studies in the emerging field of wave control through engineered wave chaos.**

The multiplexing of information carried by waves as the waves interact with a complex material (e.g., a multiply-scattering layer, a chaotic cavity, a multi-mode fiber) is a phenomenon underpinning crucial applications like communication, imaging and energy transfer[1]. Across disciplines, initial wave engineering efforts sought to compensate this information scrambling by shaping the waves incident on the medium. Examples include time-reversal focusing pioneered in acoustics[2], beam-forming for wireless communication at radiofrequencies[3], and wave-front shaping in the optical domain[4,5]. Later, it became clear that a material's complexity offers wave manipulation possibilities beyond those available in homogeneous media, leading to applications in which materials causing wave scrambling are introduced on purpose. For instance, sub-wavelength focusing in optics[6], multi-speaker listening in acoustics[7] and computational microwave imaging[8,9] are all enabled by deliberately harnessing a complex medium's property to scramble waves in a seemingly arbitrary manner.

Computational microwave imagers leverage a special case of information scrambling: multiplexing of *spatial* information with *spectral* measurement diversity – see **Figure 1**a. Waves carrying spatially encoded information about the scene are scrambled by a complex medium and the resulting field is probed at a single position at multiple independent frequencies. This circumvents the need for coherent measurements across large apertures which are notoriously difficult at radiofrequencies. The underlying mechanism is the complete mixing of degrees of freedom (DoF) in complex media irrespective of their nature (e.g., spatial, spectral, polarization) which had previously been evidenced in demonstrations of temporal focusing using spatial DoF and vice versa[10–13]. The frequency diversity of pioneering materials and devices used for multiplexing in the context of computational microwave imaging relied on metamaterial apertures[8,14] or chaotic cavities[9,15,16] with small spectral correlation lengths.

More recently, the use of engineered disordered materials rather than just seemingly arbitrary complex media is emerging in wave engineering. The disorder can be engineered either from scratch[17,18] or through the addition of a tuning mechanism to an initially random medium[19–23]. In the microwave domain, the latter was enabled largely through the emergence of tunable metasurfaces. These artificially engineered ultrathin structures can control electromagnetic wave fronts in a reconfigurable manner and are known as "tunable impedance surfaces"[24], "programmable coding metasurfaces"[25], or "spatial microwave modulators"[26]. Combining the tunable-metasurface concept with the complexity of a chaotic cavity yields what we refer to as "programmable coding metacavity" (PCM) in this work: by partially covering the cavity walls with a reconfigurable reflect-array metasurface as shown in Figure 1b, the cavity boundary conditions become programmable, thereby offering large control over the cavity wave field[26,27]. Since a sequence of random PCM coding patterns results in a series of mutually distinct fields inside the PCM, the notion of *programmable-coding* (PC) measurement diversity can be introduced.

This novel type of DoF enabled a refinement of computational microwave imaging toward single-port single-frequency operation by multiplexing spatial information across a sequence of random PCM coding patterns[28] – see Figure 1a. However, to date the deployment of PCMs has yet to reap all the benefits offered by the device's programmability. Rather than solely generating a random channel matrix, the PCM can serve as intelligent platform to custom-tailor the channel matrix properties through a judiciously chosen coding sequence. In particular, this is possible without additional hardware cost or any speed penalty during operation, the programmability being an intrinsic feature of the PCM.

In this work, we report an experimental study of custom-tailoring the space-to-PC channel matrix in a PCM through an optimized coding sequence, considering for concreteness the example of analog multiplexing of spatial information across PCM states. This specific scenario is relevant to computational imaging as well as other microwave sensing and metrology applications such as antenna-array characterization. To match key performance metrics of a channel matrix constructed from a tailored PCM coding sequence, one would have to use a significantly longer random PCM coding sequence, entailing longer acquisition times, higher power consumption and a larger processing burden. The underlying framework, however, is more general and can be applied to a number of other desirable channel matrix properties in other contexts[29] that can impact future wave engineering efforts.

In analog multiplexing, the overarching goal is to minimize the reconstruction error of the incoming spatial information. To formalize the concept, let us turn to the usual matrix formalism

$$Y = \mathbf{H}X + N, \qquad (1)$$

where the channel matrix $\mathbf{H}$ links the $n$ incoming spatially encoded pieces of information entering the PCM, $X$, to the $p$ outgoing measurements with PC diversity through which information is extracted, $Y$. $N$ denotes the measurement noise vector. As illustrated in Figure 1a, it is important to note that in such a channel matrix each row is part of a different system's scattering matrix $\mathbf{S}$ – unlike in the case of space-to-space multiplexing as encountered, for instance, in wireless communication. This has profound implications on the achievable control of the channel matrix and the optimization outcome, as will be discussed later. For the sake of generality, we do not assume any a priori knowledge about $X$ in the following; moreover, we take $\mathbf{H}$ to not be underdetermined (i.e., $p \geq n$). Then, the lowest achievable normalized mean-squared-error (NMSE) $\chi$ of a reconstruction via Tikhonov regularization[30] can be shown (see Supplementary Material) to be directly related to the singular value (SV) spectrum of the channel matrix:

$$\chi = \frac{1}{n} \sum_{i=1}^{n} \frac{1}{1 + \sigma_i^2 \rho}, \qquad (2)$$

where $\sigma_i$ is the $i$th SV of $\mathbf{H}$, $\rho = \frac{\|X\|^2}{n\zeta^2}$ is the signal-to-noise ratio (SNR) and $\zeta = \frac{\|N\|}{\sqrt{p}}$. Ideally, the SV spectrum would be as flat as possible with SV magnitudes as high as possible, in order to minimize the corruption of the reconstruction through the effect of noise on the smallest SVs. To this end, conventional methods based on random channel matrices resort to an overdetermined system ($p > n$) since the smallest normalized SV scales with $\gamma = \frac{p}{n}$ as $\frac{\sigma_{min}}{\sqrt{\frac{1}{n}\sum_{i=1}^{n}\sigma_i^2}} = 1 - \sqrt{\frac{1}{\gamma}}$.[31] However, besides being an inflexible and imprecise tool to alter the SV spectrum, using a longer random PCM coding sequence comes at the cost of more measurements. In contrast, with a judiciously chosen PCM coding sequence, the channel matrix properties can be tailored almost at wish.

Without knowledge of the SNR and the exact bounds on realizable SV distributions in a given PCM, an optimal accessible SV spectrum cannot be derived analytically from Equation 2. We hypothesize that at moderate SNR levels, maximizing the flatness of the SV spectrum will yield significant improvements of the reconstruction quality by lifting the smallest SVs above the noise level without significantly deteriorating the strongest SVs. Our focus on moderate SNR levels is justified since at very high SNR levels the smallest SVs are not significantly affected by noise anyway, while at very low SNR levels even the strongest SVs are substantially perturbed by noise. To quantify the flatness, we compute the "effective rank"[32] of the channel matrix:

$$R_{\text{eff}} = e^{-\sum_{i=1}^{n} \tilde{\sigma}_i \ln(\tilde{\sigma}_i)}, \quad (3)$$

where $\tilde{\sigma}_i = \frac{\sigma_i}{\sum_{i=1}^{n} \sigma_i}$. This metric, essentially the entropy of the SVs, should not be confused with the matrix rank: only a perfectly flat SV spectrum corresponds to $R_{\text{eff}} = n$. $R_{\text{eff}}$ is a non-integer quantity suitable for optimization.

Our experimental setup, depicted in Figure 1b and detailed in the Experimental Section, consists of a metallic cavity which contains scattering objects and two of its walls are covered with reconfigurable reflect-array metasurfaces. $n = 8$ monopole antennas inject signals that are individually modulated in-situ in phase and amplitude into the PCM while a ninth antenna probes the scrambled wave field. Given the lack of a forward model linking the coding sequence to the resulting channel matrix, we opt for an experimental iterative optimization procedure of the coding sequence as detailed in the Experimental Section.

The average downward sloping SV spectrum of a space-to-PC channel matrix based on a random coding sequence is shown in **Figure 2**a. Although the distribution of the corresponding eigenvalues of **H** in the inset appears to be roughly uniform upon visual inspection, the average effective rank $R_{\text{eff}} = 5.7 \pm 0.3$ is clearly below the value of $6.5 \pm 0.2$ expected for a random channel matrix with entries. We attribute this to the persistence of an unstirred field component in the PCM that results in a detachment of the strongest singular value from the rest (see SV distribution in Figure S5). Upon increasing $\gamma = \frac{p}{n}$ from unity to 2.5, the SV spectrum is shifted upward but retains its downward-sloping character. Increasing $\gamma$ improves the effective rank to $6.6 \pm 0.2$ and raises the "transmittance", defined as

$$T = \sum_{i=1}^{n}\sum_{j=1}^{p} |H_{i,j}|^2 = \sum_{i=1}^{n} \sigma_i^2, \qquad (4)$$

from $0.11 \pm 0.01$ to $0.28 \pm 0.03$.

Example dynamics of the iterative tailoring of the coding sequence to maximize the effective rank can be seen in Figure 2b to converge to the optimum of $R_{\text{eff}} = n = 8$ after roughly 1200 iterations. Very similar results were obtained in multiple repeats. We observe in Figure 2b that $T$ decreases slightly as we maximize $R_{\text{eff}}$, however not below the minimum value obtained with 100 random coding sequences. Indeed, the final value of $T = 0.09$ is only slightly below the average value for a random coding sequence; only the first two SVs are weaker than their random counterparts, whereas the remaining six are significantly enhanced. The final SV spectrum is flat, and the corresponding complex eigenvalues of **H** are equidistant from the origin of the complex plane.

In order to put the change in $T$ into perspective, as well as to demonstrate the generality of our approach, we also tailored a coding sequence to maximizes $T$ instead of $R_{\text{eff}}$. After 4500 iterations the optimization does not appear to have converged yet but already reached $T = 0.76$. This huge increase is largely driven by a substantial enhancement of the strongest SV – at the expense of the weaker SVs which are the most vulnerable to measurement noise. Correspondingly, the anticorrelation between $T$ and $R_{\text{eff}}$ is even more striking here: the effective rank drops to 4.2. In fact, assuming that distinct PCM configurations cannot yield exactly the same $T_j = \sum_{i=1}^{n}|H_{i,j}|^2$, it is clear that the optimal coding sequence to maximize $T$ would simply repeat $p$ times the single coding pattern that corresponds to the largest value of $T_j$ – implying that the global optimum to maximize $T$ inherently corresponds to $R_{\text{eff}} = 1$. This argument highlights a unique feature of a tailored space-to-PC channel matrix, clearly distinguishing it from a space-to-space channel matrix in a programmable environment[22]. In the context of analog multiplexing, the SV spectrum for the coding sequence tailored to maximize $T$ can be expected to yield worse results than a random coding sequence, since its smallest SV is even lower.

We now turn to the improvements in reconstruction quality enabled by tailoring the channel matrix to maximize $R_{\text{eff}}$. We inject in-situ different input vectors $X$ into the PCM and reconstruct $X$ using Tikhonov regularization based on the measurements $Y$ for a random or tailored coding sequence (see Supplementary Material). **Figure 3**a,d presents typical reconstructions for three inputs at the highest achievable experimental SNR of $\rho = 56.4$ dB. We then added synthetically more noise to the measurements to study the performance at lower SNR levels; examples for $\rho = 50$ dB and $\rho = 40$ dB are shown in Figure 3b,e and Figure 3c,f, respectively. The relation between NMSE $\chi$ and SNR $\rho$, as recovered from the in-situ measurements, is plotted in Figure 3g, in excellent agreement with the theory from Equation 2. As expected, our tailored channel matrix yields the highest enhancements in reconstruction fidelity at moderate SNR levels: $\Delta\chi = 24\%$ is achieved around $\rho = 26$ dB.

In an ideal PCM, a perfectly stirred open chaotic system with considerable loss, one would expect to obtain entries of **H** distributed as independent zero-mean Gaussian random variables (Rayleigh model).[33,34] For finite $n$, such a random matrix has non-vanishing correlations between its rows. Hence, measurements with a random coding sequence inevitably contain some redundant information that does not help to recover $X$ and consequently the effective rank of such a matrix is never full. In fact, we show in the Supplementary Material that $\langle R_{\text{eff}} \rangle = 0.8n$ for a random $n \times n$ matrix. In practice, undesired effects like an unstirred field component in our experiment yield even more correlations between different rows of **H**, adding more redundancy and ultimately resulting in an even lower effective rank. In contrast, our tailored coding sequence to maximize $R_{\text{eff}}$ ensures true orthogonality. Measurements with a tailored coding sequence do not overlap at all such that no redundant information is acquired and we can achieve the same reconstruction quality using fewer measurements.

Figure 3i compares the minimum number of measurements needed to ensure a given reconstruction quality $\chi$ at the same SNR level. In our experiment, twenty measurements ($\gamma = 2.5$) using a random coding sequence are needed to match the performance of eight measurements with a tailored coding sequence. Alternatively, we can ask what minimum SNR level is necessary using a random vs tailored coding sequence to guarantee a given reconstruction quality $\chi$. As evident in Figure 3h, to ensure, for example, $\chi \leq 5\%$ in our experiment, using eight measurements from a tailored rather than random coding sequence reduces the minimum necessary SNR from 45.6 dB to 32.4 dB. The superior characteristics of a tailored coding sequence can thus be leveraged to reduce – without any loss in performance – either the number of measurements or the minimum necessary SNR level.

Our discussion also highlights that a mathematically rigorous definition of the number of "degrees of freedom" must go beyond an integer quantity directly related to the size of **H**, and hence physical parameters $(n, p)$. Instead, it should be based on the SV spectrum of **H** using non-integer metrics like the effective rank or the eigenchannel participation number[35] that account for any intrinsic correlations. Then, the tailored coding sequence can be interpreted as increasing the number of DoF relative to that available with a random coding sequence, in our case from $5.7 \pm 0.3$ to the highest possible value of eight.

To summarize, we have demonstrated with in-situ experiments in the microwave domain that using a tailored rather than random PCM coding sequence significantly enhances the performance of analog space-to-PC information multiplexing – without any additional hardware cost or speed penalty during operation. Truly independent measurements are enabled with such a tailored coding sequence. Future work should systematically explore the range of realizable SV spectra in a given PCM, as well as other channel matrix properties, using learned[36] forward models of a PCM or models capturing the PCM's statistical behavior based on random matrix theory[37,38]. Within the more general perspective of engineered wave chaos, the use of reinforcement learning to adapt the tailored PCM coding sequence on-the-fly to a dynamically evolving application (multiplexing or other) holds great promise.

**Materials and Methods**

*Experimental Setup*: The PCM is a metallic cavity of triangular shape (base 32 cm × 32 cm, height 29 cm, quality factor ~120) containing two scattering cylinders and two of its walls are covered by reconfigurable reflect-array metasurfaces (Figure 2b). The latter consist of 152 1-bit programmable elements. At the working frequency of 5.46 GHz, each element can be programmed to mimic perfect-electric-conductor or perfect-magnetic-conductor-like boundary conditions, independently for two orthogonal polarizations.[39] The working principle is based on the hybridization of resonances, the switching mechanism relies on PIN diodes.[40] A signal, generated by a vector network analyzer, is split by an eight-way power divider and each way is individually modulated in amplitude and phase by an IQ modulator before being injected into the PCM via monopole antennas placed randomly on the PCM's surface (Figure 2c). A ninth such antenna probes the field inside the PCM and is connected to the vector network analyzer's receiving port. Power splitting, losses in the IQ modulators and absorption in the PCM explain the low values of $T$. To test the multiplexing performance, we inject in-situ a series of 25 different input vectors $X$ and measure the corresponding outputs $Y$. We compare the performance of the custom-tailored channel matrix with that of 25 channel matrices corresponding to 25 different random coding sequences.

*Optimization Algorithm*: To identify a suitable coding sequence for the PCM of length $p = n$ that optimizes our chosen metric $R_{\text{eff}}$, we use an iterative experimental optimization procedure. First, we evaluate the metric for 100 random coding sequences and select the sequence yielding the highest $R_{\text{eff}}$. Then, we iteratively refine that coding sequence: for every iteration, we randomly select $z$ elements and check if flipping their configuration is advantageous regarding our metric in which case we update the coding sequence accordingly. We gradually reduce $z$ as the iteration index $w$ increases: $z = \max(\text{int}(0.97^w \frac{304}{2}), 1)$. We stop the optimization after it saturates.

**Acknowledgements**

The authors are supported by the French "Agence Nationale de la Recherche" under reference ANR-17-ASTR-0017. The metasurface prototypes were purchased from Greenerwave. The authors acknowledge fruitful discussions with Olivier Legrand, Fabrice Mortessagne and Dmitry Savin.

The project was initiated, conceptualized and conducted by P.d.H. The derivation of Equation 2 was proposed by M.D. U.K. provided the experimental equipment and contributed to the design of the experimental setup and the in-situ realization. All authors thoroughly discussed the results. The manuscript was written by P.d.H. and reviewed by all authors.

The authors declare no conflict of interests.

**References**

[1] P. Sebbah, *Waves and Imaging through Complex Media*, Springer Science & Business Media, **2001**.
[2] M. Fink, *Phys. Today* **1997**, *50*, 34.
[3] E. Telatar, *Eur. Trans. Telecomm.* **1999**, *10*, 585.
[4] I. M. Vellekoop, A. P. Mosk, *Opt. Lett.* **2007**, *32*, 2309.
[5] S. Rotter, S. Gigan, *Rev. Mod. Phys.* **2017**, *89*, 015005.
[6] I. M. Vellekoop, A. Lagendijk, A. P. Mosk, *Nat. Photon.* **2010**, *4*, 320.
[7] Y. Xie, T.-H. Tsai, A. Konneker, B.-I. Popa, D. J. Brady, S. A. Cummer, *Proc. Natl. Acad. Sci. USA* **2015**, *112*, 10595.
[8] J. Hunt, T. Driscoll, A. Mrozack, G. Lipworth, M. Reynolds, D. Brady, D. R. Smith, *Science* **2013**, *339*, 310.
[9] T. Fromenteze, O. Yurduseven, M. F. Imani, J. Gollub, C. Decroze, D. Carsenat, D. R. Smith, *Appl. Phys. Lett.* **2015**, *106*, 194104.
[10] F. Lemoult, G. Lerosey, J. de Rosny, M. Fink, *Phys. Rev. Lett.* **2009**, *103*, 173902.
[11] J. Aulbach, B. Gjonaj, P. M. Johnson, A. P. Mosk, A. Lagendijk, *Phys. Rev. Lett.* **2011**, *106*, 103901.
[12] O. Katz, E. Small, Y. Bromberg, Y. Silberberg, *Nat. Photon.* **2011**, *5*, 372.
[13] D. J. McCabe, A. Tajalli, D. R. Austin, P. Bondareff, I. A. Walmsley, S. Gigan, B. Chatel, *Nat. Commun.* **2011**, *2*, 447.
[14] J. N. Gollub, O. Yurduseven, K. P. Trofatter, D. Arnitz, M. F. Imani, T. Sleasman, M. Boyarsky, A. Rose, A. Pedross-Engel, H. Odabasi, T. Zvolensky, G. Lipworth, D. Brady, D. L. Marks, M. S. Reynolds, D. R. Smith, *Sci. Rep.* **2017**, *7*, 42650.
[15] T. Fromenteze, C. Decroze, D. Carsenat, *IEEE Trans. Antennas Propag.* **2015**, *63*, 593.
[16] T. Fromenteze, E. L. Kpre, D. Carsenat, C. Decroze, T. Sakamoto, *IEEE Access* **2016**, *4*, 1050.
[17] L. S. Froufe-Pérez, M. Engel, J. J. Sáenz, F. Scheffold, *Proc. Natl. Acad. Sci. USA* **2017**, *114*, 9570.
[18] M. Jang, Y. Horie, A. Shibukawa, J. Brake, Y. Liu, S. M. Kamali, A. Arbabi, H. Ruan, A. Faraon, C. Yang, *Nat. Photon.* **2018**, *12*, 84.
[19] K. G. Makris, A. Brandstötter, P. Ambichl, Z. H. Musslimani, S. Rotter, *Light Sci. Appl.* **2017**, *6*, e17035.
[20] E. Rivet, A. Brandstötter, K. G. Makris, H. Lissek, S. Rotter, R. Fleury, *Nat. Phys.* **2018**, *14*, 942.
[21] P. del Hougne, G. Lerosey, *Phys. Rev. X* **2018**, *8*, 041037.
[22] P. del Hougne, M. Fink, G. Lerosey, *Nat. Electron.* **2019**, *2*, 36.
[23] S. Resisi, Y. Viernik, S. Popoff, Y. Bromberg, *arXiv:1910.02798* **2019**.

[24] D. F. Sievenpiper, J. H. Schaffner, H. J. Song, R. Y. Loo, G. Tangonan, *IEEE Trans. Antennas Propagat.* **2003**, *51*, 2713.
[25] T. J. Cui, M. Q. Qi, X. Wan, J. Zhao, Q. Cheng, *Light Sci. Appl.* **2014**, *3*, e218.
[26] M. Dupré, P. del Hougne, M. Fink, F. Lemoult, G. Lerosey, *Phys. Rev. Lett.* **2015**, *115*, 017701.
[27] P. del Hougne, F. Lemoult, M. Fink, G. Lerosey, *Phys. Rev. Lett.* **2016**, *117*, 134302.
[28] T. Sleasman, M. F. Imani, J. N. Gollub, D. R. Smith, *Phys. Rev. Applied* **2016**, *6*, 054019.
[29] N. Bender, H. Yılmaz, Y. Bromberg, H. Cao, *APL Photonics* **2019**, *4*, 110806.
[30] A. N. Tikhonov, *Soviet Math. Dokl.* **1963**, *4*, 1035.
[31] S. Popoff, G. Lerosey, M. Fink, A. C. Boccara, S. Gigan, *Nat. Commun.* **2010**, *1*, 81.
[32] O. Roy, M. Vetterli, *15th European Signal Processing Conference* **2007**, 606.
[33] J.-H. Yeh, T. M. Antonsen, E. Ott, S. M. Anlage, *Phys. Rev. E* **2012**, *85*, 015202.
[34] S. Kumar, A. Nock, H.-J. Sommers, T. Guhr, B. Dietz, M. Miski-Oglu, A. Richter, F. Schäfer, *Phys. Rev. Lett.* **2013**, *111*, 030403.
[35] M. Davy, Z. Shi, A. Z. Genack, *Phys. Rev. B* **2012**, *85*, 035105.
[36] S. Ma, B. Xiao, R. Hong, B. Addissie, Z. Drikas, T. Antonsen, E. Ott, S. Anlage, *arXiv:1908.04716* **2019**.
[37] U. Kuhl, O. Legrand, F. Mortessagne, *Fortschr. Phys.* **2013**, *61*, 404.
[38] G. Gradoni, J.-H. Yeh, B. Xiao, T. M. Antonsen, S. M. Anlage, E. Ott, *Wave Motion* **2014**, *51*, 606.
[39] P. del Hougne, Shaping Green's Functions in Cavities with Tunable Boundary Conditions: From Fundamental Science to Applications, Université Sorbonne Paris Cité, **2018**.
[40] N. Kaina, M. Dupré, M. Fink, G. Lerosey, *Opt. Express* **2014**, *22*, 18881.

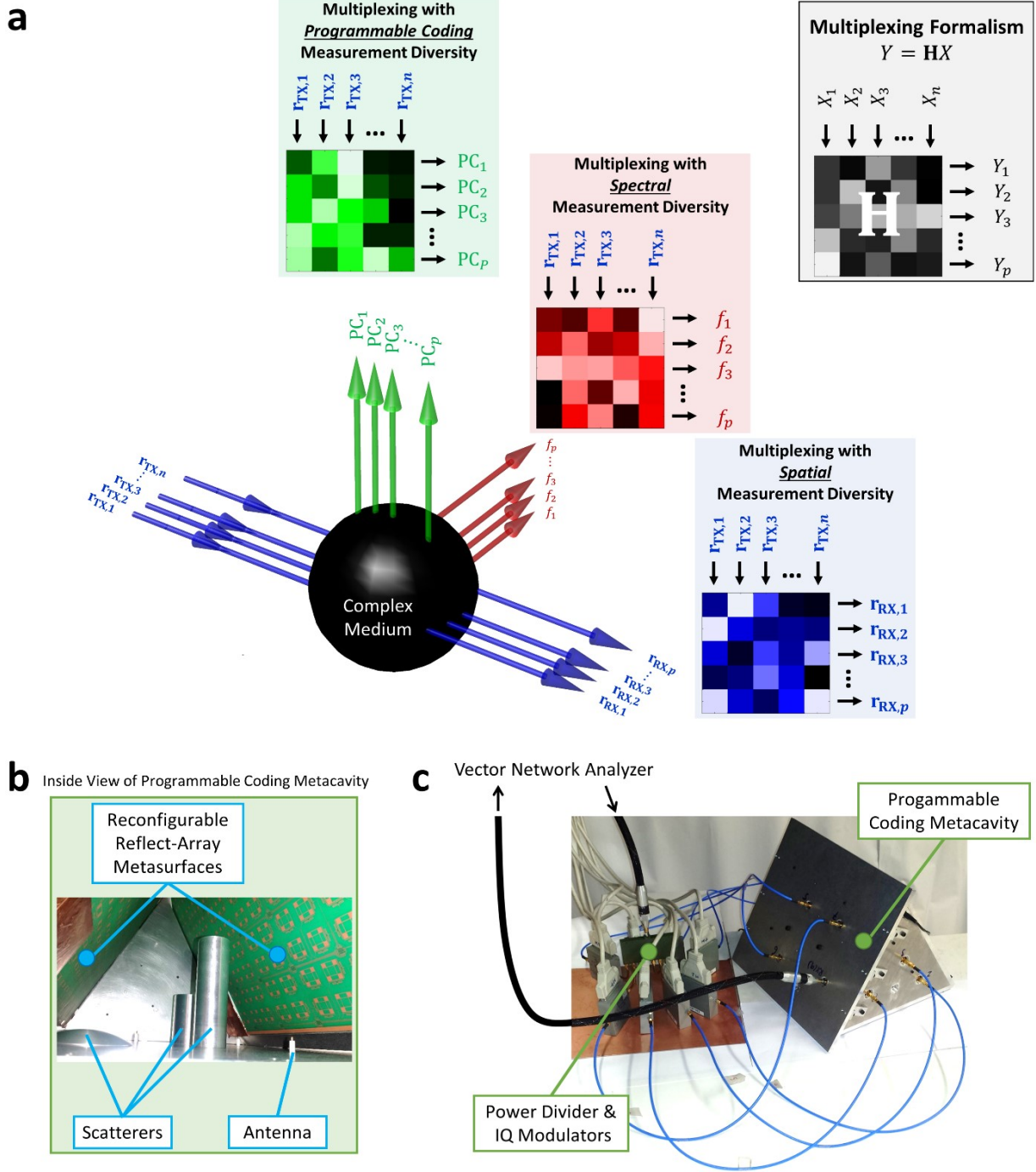

**Figure 1. a, Principle of information multiplexing with different types of measurement diversity.** A complex medium scrambles $n$ pieces of spatially ($\mathbf{r}_{\mathbf{TX},i}$, blue) encoded information. To recover these inputs, the scrambled wave field is probed with $p$ independent measurements which can be of spatial nature ($\mathbf{r}_{\mathbf{RX},i}$, blue), spectral nature ($f_i$, red) or correspond to different configurations of the complex medium ($PC_i$, green). **b, Programmable coding metacavity (PCM).** The PCM is an irregular metallic cavity containing scattering objects and two walls are covered by reconfigurable reflect-array metasurfaces. **c, In-situ setup.** Eight signals, modulated in-situ by eight IQ modulators, excite the cavity via eight antennas. A ninth antenna probes the resulting cavity wave field for different states of the PCM.

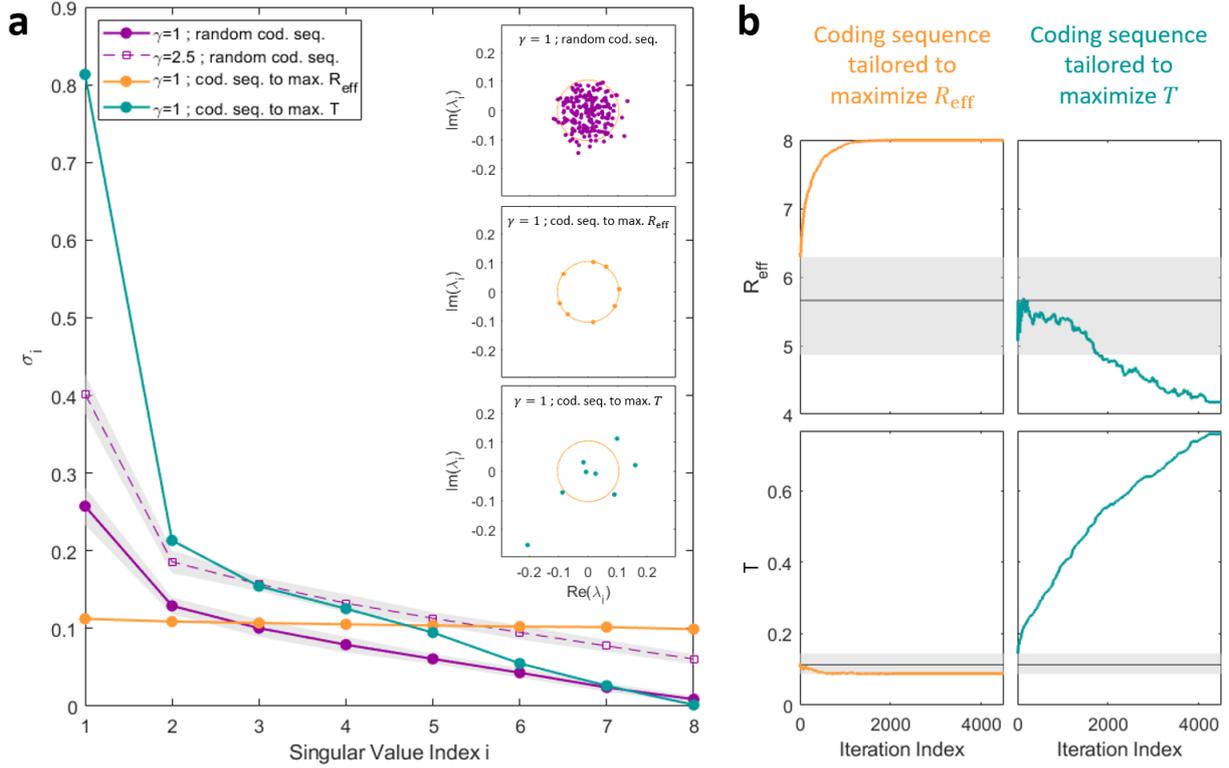

**Figure 2. Dynamics and outcome of tailoring the channel matrix. a,** Singular value distributions for random channel matrices with $\gamma = 1$ and $\gamma = 2.5$ (averaged over 25 realizations), as well as for a channel matrix with $\gamma = 1$ tailored to maximize $R_{\text{eff}}$ or $T$, respectively. The shaded areas indicate the standard deviations for the cases with random coding sequences. The insets indicate the distribution of the eigenvalues $\lambda_i$ of **H** in the complex plane for 25 random channel matrices with $\gamma = 1$ (top), for the channel matrix tailored to maximize $R_{\text{eff}}$ (middle) and for the channel matrix tailored to maximize $T$ (bottom). For reference, an orange circle is indicated whose radius is the average magnitude of the eigenvalues for the channel matrix tailored to maximize $R_{\text{eff}}$. **b,** Evolution of $R_{\text{eff}}$ (top row) and $T$ (bottom row) over the course of the iterative optimization of the coding sequence to maximize $R_{\text{eff}}$ (left column) or $T$ (right column). The shaded area and continuous black line indicate the range of values of $R_{\text{eff}}$ and $T$ spanned over 100 random realizations and their average, respectively.

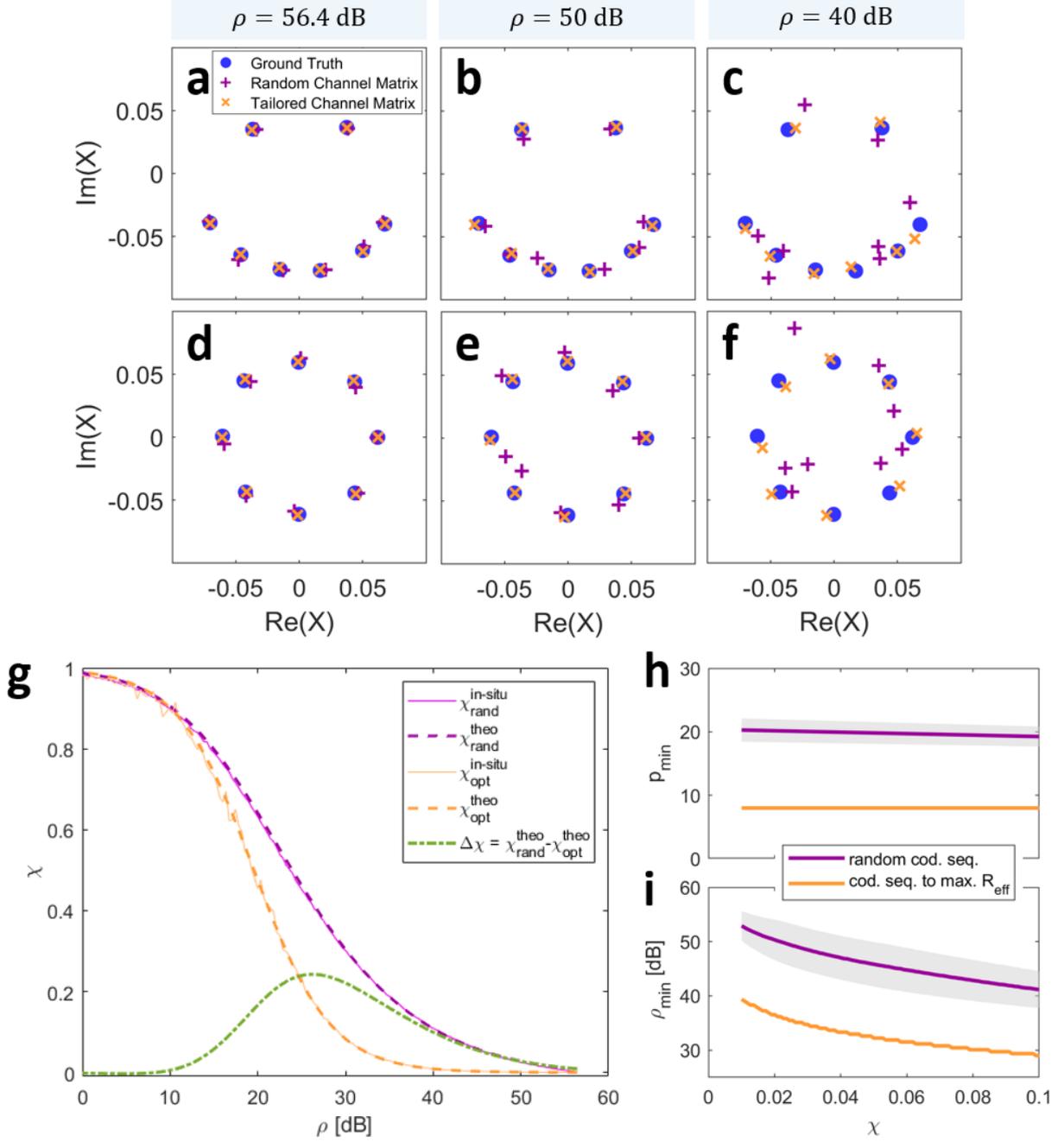

**Figure 3. Reconstruction quality with random (purple) vs. tailored (orange) channel matrix. a-f,** Sample reconstructions of two injected signals (smiley, circle) at three distinct SNR levels: 56.4 dB (left), 50.0 dB (middle), 40.0 dB (right). $\rho = 56.4$ dB is the highest realizable SNR level in our experiment. The displayed realizations have been chosen such that their reconstruction error is closest to the average reconstruction error. **g,** Dependence of the average normalized mean squared reconstruction error, $\chi$, on the SNR $\rho$. The corresponding curves for a reconstruction based on in-situ measurements (continuous lines) are contrasted with the curves dictated by the theory from Equation 2 (dashed lines). The difference between the two curves for random vs tailored coding sequences is plotted in green. **h,** Minimum number of measurements $p_{min}$ necessary to guarantee a given reconstruction quality $\chi$ at the SNR level at which $p = n$ guarantees this reconstruction quality using the tailored coding sequence. The shaded area indicates the standard deviation when using random coding sequences. **i,** Lowest necessary SNR level $\rho_{min}$ that guarantees a given reconstruction quality $\chi$ for a system with $\gamma = 1$ for the case of using a random or tailored coding sequence.

# Supplementary Material

# 1 Notation

| | |
|---|---|
| $n$ | number of incoming channels |
| $p$ | number of outgoing channels, we assume in the following that $p \geq n$ |
| $X$ | input vector ($n \times 1$) |
| $Y$ | output vector ($p \times 1$) |
| $\mathbf{H}$ | channel matrix ($p \times n$) |
| $N$ | measurement noise vector ($p \times 1$) |
| $\tilde{X}$ | reconstructed input vector ($n \times 1$) |
| $\mathcal{H}$ | reconstruction operator ($n \times p$) |
| $\varepsilon$ | reconstruction error: $\varepsilon = \|X - \tilde{X}\|$ |
| $\zeta$ | standard deviation of Gaussian measurement noise: $\zeta = \|N\|/\sqrt{p}$ |
| $\dagger$ | denotes the conjugate transpose |
| $\|...\|$ | denotes the 2-norm |
| $\langle ... \rangle$ | denotes the average |

# 2 Link between Matrix and SVD Versions of Tikhonov Regularization

The following derivation of a well-known result is provided for the sake of completeness.

The matrix version of Tikhonov regularization[1,2] with Tikhonov parameter $\alpha$ reads:
$$\mathcal{H} = (\mathbf{H}^\dagger \mathbf{H} + \alpha \mathbf{I})^{-1} \mathbf{H}^\dagger$$

We now use the singular value decomposition $\mathbf{H} = \mathbf{U\Sigma V}^\dagger$.
If $p = n$, $\mathbf{U}$, $\mathbf{\Sigma}$ and $\mathbf{V}$ are $n \times n$ matrices.
If $p > n$, there are at most $n$ non-zero singular values such that $\mathbf{U}$ is a $p \times n$ matrix while $\mathbf{\Sigma}$ and $\mathbf{V}$ are $n \times n$ matrices.
Using $\mathbf{H} = \mathbf{U\Sigma V}^\dagger$,
$$\mathcal{H} = ((\mathbf{U\Sigma V}^\dagger)^\dagger (\mathbf{U\Sigma V}^\dagger) + \alpha \mathbf{I})^{-1} (\mathbf{U\Sigma V}^\dagger)^\dagger$$

Using the properties $(\mathbf{A_1 A_2 ... A_n})^\dagger = \mathbf{A_n^\dagger ... A_2^\dagger A_1^\dagger}$ and $(\mathbf{A}^\dagger)^\dagger = \mathbf{A}$,
$$\mathcal{H} = ((\mathbf{V\Sigma}^\dagger \mathbf{U}^\dagger)(\mathbf{U\Sigma V}^\dagger) + \alpha \mathbf{I})^{-1} (\mathbf{V\Sigma}^\dagger \mathbf{U}^\dagger)$$

Using the unitarity property of $\mathbf{V}$ and $\mathbf{U}$, we can simplify $\mathbf{U}^\dagger \mathbf{U} = \mathbf{I}$ and write $\mathbf{I} = \mathbf{V I V}^\dagger$, yielding
$$\mathcal{H} = (\mathbf{V\Sigma}^\dagger \mathbf{\Sigma V}^\dagger + \alpha \mathbf{V I V}^\dagger)^{-1} (\mathbf{V\Sigma}^\dagger \mathbf{U}^\dagger)$$
$$\mathcal{H} = (\mathbf{V}(\mathbf{\Sigma}^\dagger \mathbf{\Sigma} + \alpha \mathbf{I})\mathbf{V}^\dagger)^{-1} (\mathbf{V\Sigma}^\dagger \mathbf{U}^\dagger)$$

Using the property $(\mathbf{A_1 A_2 ... A_n})^{-1} = \mathbf{A_n^{-1} ... A_2^{-1} A_1^{-1}}$,
$$\mathcal{H} = ((\mathbf{V}^\dagger)^{-1} (\mathbf{\Sigma}^\dagger \mathbf{\Sigma} + \alpha \mathbf{I})^{-1} \mathbf{V}^{-1})(\mathbf{V\Sigma}^\dagger \mathbf{U}^\dagger)$$

Using the unitarity property of $\mathbf{V}$, we can simplify $(\mathbf{V}^\dagger)^{-1} = \mathbf{V}$ and $\mathbf{V}^{-1}\mathbf{V} = \mathbf{I}$, yielding
$$\mathcal{H} = \mathbf{V}(\mathbf{\Sigma}^\dagger \mathbf{\Sigma} + \alpha \mathbf{I})^{-1} \mathbf{\Sigma}^\dagger \mathbf{U}^\dagger$$

Using $\Sigma = \text{diag}(\sigma_i)$ and $\alpha\mathbf{I} = \text{diag}(\alpha)$, and the property that $(\text{diag}(d_i))^{-1} = \text{diag}\left(\frac{1}{d_i}\right)$,

$$\mathcal{H} = \mathbf{V}\,\text{diag}\left(\frac{\sigma_i}{\sigma_i^2 + \alpha}\right)\mathbf{U}^\dagger$$

## 3   Link between Reconstruction Error and Singular Value Distribution

The aim of this section is to link the reconstruction error $\langle\varepsilon\rangle = \langle\|\mathcal{H}Y - X\|\rangle$ to the singular value distribution of a given channel matrix $\mathbf{H}$. The average here refers to different realizations of $X$ and $N$.

Consider a generalized reconstruction method:

$$\mathcal{H} = \mathbf{V}\,\text{diag}(w(\sigma_i))\mathbf{U}^\dagger = \sum_{i=1}^{n} \mathbf{v}_i\, w(\sigma_i)\, \mathbf{u}_i^\dagger$$

The function $w(\sigma_i)$ takes the following form for a few well-known reconstruction methods:
- Pseudo-Inversion: $\quad w(\sigma_i) = \frac{1}{\sigma_i}$
- Truncated–SVD Inversion: $\quad w(\sigma_i) = \begin{cases} \frac{1}{\sigma_i} & \text{for } \sigma_i > \alpha \\ 0 & \text{for } \sigma_i < \alpha \end{cases}$
- Tikhonov Regularization: $\quad w(\sigma_i) = \frac{\sigma_i}{\sigma_i^2 + \alpha}$

$\alpha$ is a regularization parameter whose value is to be determined (see below).

$$\begin{aligned}
\mathcal{H}Y &= \mathcal{H}(\mathbf{H}X + N) \\
&= \left(\mathbf{V}\,\text{diag}(w(\sigma_i))\mathbf{U}^\dagger\right)\left((\mathbf{U}\,\text{diag}(\sigma_i)\mathbf{V}^\dagger)X + N\right) \\
&= \mathbf{V}\,\text{diag}(w(\sigma_i))\mathbf{U}^\dagger\mathbf{U}\,\text{diag}(\sigma_i)\mathbf{V}^\dagger X + \mathbf{V}\,\text{diag}(w(\sigma_i))\mathbf{U}^\dagger N \\
&= \mathbf{V}\,\text{diag}(w(\sigma_i)\sigma_i)\mathbf{V}^\dagger X + \mathbf{V}\,\text{diag}(w(\sigma_i))\mathbf{U}^\dagger N
\end{aligned}$$

In combination with $X = \mathbf{V}\mathbf{V}^\dagger X$, this yields
$$\begin{aligned}
\varepsilon &= \|\mathcal{H}Y - X\| \\
&= \|\mathbf{V}\,\text{diag}(w(\sigma_i)\sigma_i)\mathbf{V}^\dagger X + \mathbf{V}\,\text{diag}(w(\sigma_i))\mathbf{U}^\dagger N - \mathbf{V}\mathbf{V}^\dagger X\| \\
&= \|\mathbf{V}\,\text{diag}(w(\sigma_i)\sigma_i - 1)\mathbf{V}^\dagger X + \mathbf{V}\,\text{diag}(w(\sigma_i))\mathbf{U}^\dagger N\| \\
&= \|\mathbf{A}X + \mathbf{B}N\|
\end{aligned}$$

We assume that the measurement noise vector $N$ and the signal vector $X$ are uncorrelated. Thus, the general property $\|P + Q\|^2 = \|P\|^2 + \|Q\|^2 + 2PQ$ simplifies to $\|P + Q\|^2 = \|P\|^2 + \|Q\|^2$ because $\langle PQ\rangle = 0$ if $P$ and $Q$ are independent random complex vectors. We hence find

$$\langle\varepsilon^2\rangle = \langle\|\mathbf{A}X + \mathbf{B}N\|\rangle = \langle\|\mathbf{A}X\|^2\rangle + \langle\|\mathbf{B}N\|^2\rangle + 2\langle\mathbf{A}X\mathbf{B}N\rangle = \langle\|\mathbf{A}X\|^2\rangle + \langle\|\mathbf{B}N\|^2\rangle$$

Incidentally, $\langle\varepsilon^2\rangle$ is the mean-squared-error (MSE) of the reconstruction.

Let us now evaluate each term in turn.

$$\langle\|\mathbf{A}X\|^2\rangle = \langle(\mathbf{A}X)^\dagger\mathbf{A}X\rangle = \langle X^\dagger\mathbf{A}^\dagger\mathbf{A}X\rangle$$

$$\begin{aligned}
\mathbf{A}^\dagger \mathbf{A} &= (\mathbf{V}\,\mathrm{diag}(w(\sigma_i)\sigma_i - 1)\mathbf{V}^\dagger)^\dagger (\mathbf{V}\,\mathrm{diag}(w(\sigma_i)\sigma_i - 1)\mathbf{V}^\dagger) \\
&= \left((\mathbf{V}^\dagger)^\dagger \left(\mathrm{diag}(w(\sigma_i)\sigma_i - 1)\right)^\dagger \mathbf{V}^\dagger\right)(\mathbf{V}\,\mathrm{diag}(w(\sigma_i)\sigma_i - 1)\mathbf{V}^\dagger) \\
&= \mathbf{V}\,\mathrm{diag}((w(\sigma_i)\sigma_i - 1)^2)\,\mathbf{V}^\dagger
\end{aligned}$$

Hence,
$$\begin{aligned}
\langle \|\mathbf{A}X\|^2 \rangle &= \langle X^\dagger \mathbf{A}^\dagger \mathbf{A} X \rangle \\
&= \langle X^\dagger \mathbf{V}\,\mathrm{diag}((w(\sigma_i)\sigma_i - 1)^2)\,\mathbf{V}^\dagger X \rangle \\
&= \langle X^\dagger \left[\sum_{i=1}^{n} v_i (w(\sigma_i)\sigma_i - 1)^2 v_i^\dagger\right] X \rangle
\end{aligned}$$

<span style="color:blue">Since the singular value distribution is fixed,</span>
$$\begin{aligned}
&= \sum_{i=1}^{n} (w(\sigma_i)\sigma_i - 1)^2 \langle X^\dagger v_i v_i^\dagger X \rangle \\
&= \sum_{i=1}^{n} (w(\sigma_i)\sigma_i - 1)^2 \langle |X^\dagger v_i|^2 \rangle \\
&= \sum_{i=1}^{n} (w(\sigma_i)\sigma_i - 1)^2 \langle \left|\sum_{j=1}^{n} X_j^* v_{i,j}\right|^2 \rangle
\end{aligned}$$

<span style="color:blue">Since $X_j^*$ is statistically independent from $v_{i,j}$,</span>
$$= \sum_{i=1}^{n} (w(\sigma_i)\sigma_i - 1)^2 \langle \sum_{j=1}^{n} |X_j^*|^2 |v_{i,j}|^2 \rangle$$

<span style="color:blue">By definition, $\|v_i\| = 1$</span>
$$\begin{aligned}
&= \sum_{i=1}^{n} (w(\sigma_i)\sigma_i - 1)^2 \frac{\|X\|^2}{n} \\
&= \frac{\|X\|^2}{n} \sum_{i=1}^{n} (w(\sigma_i)\sigma_i - 1)^2
\end{aligned}$$

Similarly, we evaluate the second term.
$$\begin{aligned}
\langle \|\mathbf{B}N\|^2 \rangle &= \langle N^\dagger \mathbf{B}^\dagger \mathbf{B} N \rangle \\
&= \langle N^\dagger (\mathbf{V}\,\mathrm{diag}(w(\sigma_i))\mathbf{U}^\dagger)^\dagger (\mathbf{V}\,\mathrm{diag}(w(\sigma_i))\mathbf{U}^\dagger) N \rangle \\
&= \langle N^\dagger \mathbf{U}\,\mathrm{diag}(w(\sigma_i))\mathbf{V}^\dagger \mathbf{V}\,\mathrm{diag}(w(\sigma_i))\mathbf{U}^\dagger N \rangle \\
&= \langle N^\dagger \mathbf{U}\,\mathrm{diag}\left((w(\sigma_i))^2\right)\mathbf{U}^\dagger N \rangle \\
&= \langle N^\dagger \left[\sum_{i=1}^{n} u_i (w(\sigma_i))^2 u_i^\dagger\right] N \rangle \\
&= \sum_{i=1}^{n} (w(\sigma_i))^2 \langle N^\dagger u_i u_i^\dagger N \rangle
\end{aligned}$$

$$\begin{aligned}
&= \sum_{i=1}^{n}(w(\sigma_i))^2 \langle |N^\dagger u_i|^2 \rangle \\
&= \sum_{i=1}^{n}(w(\sigma_i))^2 \langle \left|\sum_{j=1}^{p} N_j^\dagger u_{i,j}\right|^2 \rangle \\
&= \sum_{i=1}^{n}(w(\sigma_i))^2 \langle \sum_{j=1}^{p} |N_j^\dagger|^2 |u_{i,j}|^2 \rangle \\
&= \sum_{i=1}^{n}(w(\sigma_i))^2 \frac{\|N\|^2}{p} \\
&= \zeta^2 \sum_{i=1}^{n}(w(\sigma_i))^2
\end{aligned}$$

Thus, putting everything together,

$$\langle \varepsilon^2 \rangle = \langle \|AX\|^2 \rangle + \langle \|BN\|^2 \rangle$$

$$= \left[\frac{\|X\|^2}{n}\sum_{i=1}^{n}(w(\sigma_i)\sigma_i - 1)^2\right] + \left[\zeta^2 \sum_{i=1}^{n}(w(\sigma_i))^2\right]$$

## 4  Optimal Value of Tikhonov Parameter and Lowest Achievable Reconstruction Error

The optimal value of the Tikhonov parameter, $\alpha_{opt}$, minimizes $\langle \varepsilon^2 \rangle$. Let us first insert the expression for $w(\sigma_i)$ for Tikhonov regularization into the general expression for $\langle \varepsilon^2 \rangle$ derived above.

$$\begin{aligned}
\langle \varepsilon^2 \rangle &= \left[\frac{\|X\|^2}{n}\sum_{i=1}^{n}(w(\sigma_i)\sigma_i - 1)^2\right] + \left[\zeta^2 \sum_{i=1}^{n}(w(\sigma_i))^2\right] \\
&= \left[\frac{\|X\|^2}{n}\sum_{i=1}^{n}\left(\frac{\sigma_i}{\sigma_i^2 + \alpha}\sigma_i - 1\right)^2\right] + \left[\zeta^2 \sum_{i=1}^{n}\left(\frac{\sigma_i}{\sigma_i^2 + \alpha}\right)^2\right] \\
&= \left[\frac{\|X\|^2}{n}\sum_{i=1}^{n}\frac{\alpha^2}{(\sigma_i^2 + \alpha)^2}\right] + \left[\zeta^2 \sum_{i=1}^{n}\frac{\sigma_i^2}{(\sigma_i^2 + \alpha)^2}\right]
\end{aligned}$$

### 4.1  Channel Matrix with $R_{\text{eff}} = n$

For a channel matrix with $R_{\text{eff}} = n$, all singular values are identical. Thus,

$$\langle \varepsilon^2 \rangle = \frac{\|X\|^2}{n}\frac{n\alpha^2}{(\sigma_0^2 + \alpha)^2} + \zeta^2 \frac{n\sigma_0^2}{(\sigma_0^2 + \alpha)^2}$$

$$= \frac{\|X\|^2 \alpha^2 + \zeta^2 n \sigma_0^2}{(\sigma_0^2 + \alpha)^2}$$

In order to find the value of $\alpha$ that minimizes $\langle \varepsilon^2 \rangle$, we first compute the following derivative:

$$\frac{\partial \langle \varepsilon^2 \rangle}{\partial \alpha} = \frac{2\sigma_0^2 \left( \alpha \|X\|^2 - n\zeta^2 \right)}{(\alpha + \sigma_0^2)^3}$$

Then, using $\left. \frac{\partial \langle \varepsilon^2 \rangle}{\partial \alpha} \right|_{\alpha = \alpha_{opt}} = 0$ yields $\alpha_{opt}\|X\|^2 - n\zeta^2 = 0$ such that

$$\boxed{\alpha_{opt} = \frac{n}{\|X\|^2} \zeta^2 = \frac{1}{\rho},} \qquad (S1)$$

where $\rho = \frac{\|X\|^2}{\|N\|^2} = \frac{\|X\|^2}{n\zeta^2}$ is the Signal-to-Noise Ratio (SNR).

We verified Equation S1 using our experimental data from in-situ transmissions with the channel matrix tailored to maximize the effective rank.

First, we determined the experimental noise level. To this end, for a series of input vectors $X$ injected in-situ and the corresponding measurements $Y$, we computed $N = Y - \mathbf{H}X$. Based on multiple realizations of $\mathbf{H}$ and $X$, we then studied the distribution of the entries of $N$ which were observed to follow a Gaussian distribution with zero mean and standard deviation $\zeta_{exp}$. The highest achievable experimental SNR level was thus evaluated to be $\rho = \frac{\|X\|^2}{n\zeta_{exp}^2} = 56.4$ dB. Lower SNR levels were emulated by numerically adding white noise of standard deviation $\zeta_{synth}$ to the measured data, yielding an overall noise standard deviation of $\zeta = \sqrt{\zeta_{exp}^2 + \zeta_{synth}^2}$.

Second, using the experimentally obtained channel matrix with optimized effective rank, for different levels of SNR (controlled by numerically adding noise with zero mean and standard deviation $\zeta_{synth}$ to the measured values of $Y$), we tried for each SNR level a wide range of possible values of $\alpha$ to identify heuristically the one that yielded the lowest $\langle \varepsilon^2 \rangle$. The identified values (continuous line) are compared with those predicted by the above derivation (dashed line) and show an excellent agreement in Figure S1. Note that the optimal values of $\alpha$ are chosen from a logarithmically spaced list of 2000 points (between $10^{-12}$ and 10). Similarly, the considered values of $\zeta_{synth}$ are a logarithmically spaced list of 500 points (between $10^{-10}$ and $10^{-1}$). This explains why more fluctuations are seen in Figure S1 for larger $\zeta^2$. The results are averaged over 25 realizations of the synthetically added noise.

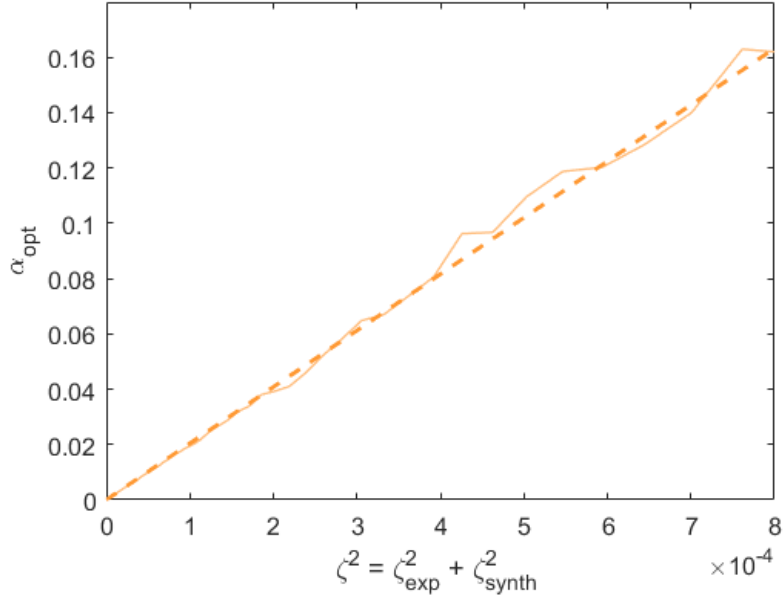

**Figure S1.** *Comparison of empirically identified and predicted optimal value $\alpha_{opt}$ of the Tikhonov parameter for different SNR levels. The optimal heuristically identified value is shown as continuous line, the value predicted by Equation S1 is shown as dashed line.*

Given the optimal value of the Tikhonov parameter, we can proceed to determine the lowest achievable reconstruction error.

The lowest achievable $\langle \varepsilon^2 \rangle$ is hence $\frac{\|X\|^2 \alpha_{opt}^2 + \zeta^2 n \sigma_0^2}{(\sigma_0^2 + \alpha_{opt})^2} = \frac{\|X\|^2 \left(\frac{n}{\|X\|^2}\zeta^2\right)^2 + \zeta^2 n \sigma_0^2}{\left(\sigma_0^2 + \frac{n}{\|X\|^2}\zeta^2\right)^2} = \|X\|^2 \frac{1}{1 + \frac{\sigma_0^2 \|X\|^2}{n\zeta^2}}$.

Finally, this yields the lowest achievable normalized mean-squared-error $\chi$:

$$\chi = \frac{\langle \varepsilon^2 \rangle|_{\alpha=\alpha_{opt}}}{\|X\|^2} = \frac{1}{1 + \frac{\sigma_0^2 \|X\|^2}{n\zeta^2}} = \frac{1}{1 + \sigma_0^2 \rho} \ . \quad (S2)$$

In the limit of $\zeta \to 0$, $\frac{\langle \varepsilon^2 \rangle}{\|X\|^2} \to \frac{n}{\sigma_0^2 \|X\|^2} \zeta^2$. This quadratic behavior is clearly observed in Figure S2. In the limit of $\zeta \to \infty$, $\frac{\langle \varepsilon^2 \rangle}{\|X\|^2} \to 1$, i.e. the reconstructed signal is 100% wrong on average, as expected.

We use 15 out of the 25 inputs injected in-situ to estimate the optimal value $\alpha_{opt}$ of the Tikhonov parameter and then test the reconstruction quality on the "unseen" inputs. As shown in Figure S2, we obtain an excellent agreement with theoretical predictions based on Equation S2.

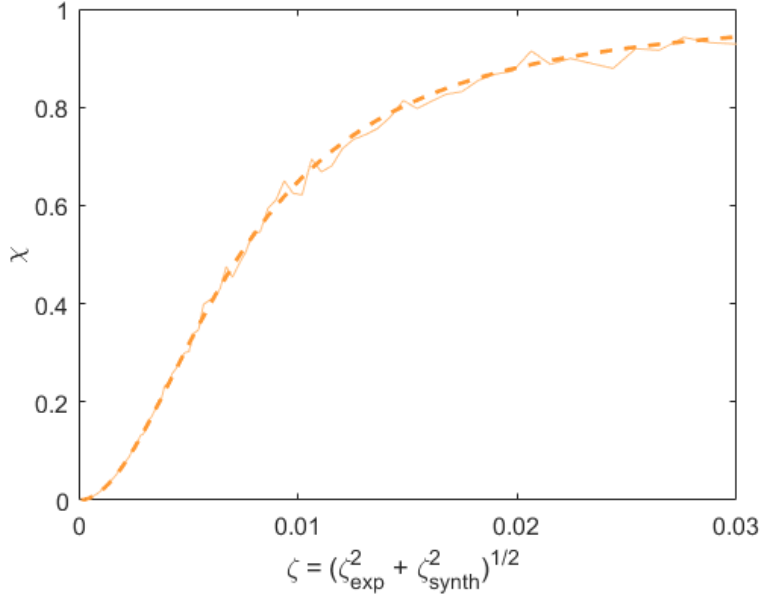

**Figure S2.** *Comparison of the lowest achievable normalized mean squared error, χ, as a function of the standard deviation ζ of the measurement noise. The continuous line represents the reconstruction error on unseen data using, for a given level of noise, the previously determined optimal Tikhonov parameter value (see Figure S1). The dashed line corresponds to the theoretical prediction from Equation S2.*

### 4.2  General Channel Matrix

Using $\langle \varepsilon^2 \rangle = \left[ \frac{\|X\|^2}{n} \sum_{i=1}^{n} \frac{\alpha^2}{(\sigma_i^2 + \alpha)^2} \right] + \left[ \zeta^2 \sum_{i=1}^{n} \frac{\sigma_i^2}{(\sigma_i^2 + \alpha)^2} \right]$, we find

$$\frac{\partial \langle \varepsilon^2 \rangle}{\partial \alpha} = \left[ \frac{\|X\|^2}{n} \sum_{i=1}^{n} \frac{2\alpha \sigma_i^2}{(\sigma_i^2 + \alpha)^3} \right] + \left[ \zeta^2 \sum_{i=1}^{n} \frac{-2\sigma_i^2}{(\sigma_i^2 + \alpha)^3} \right]$$

$$= \sum_{i=1}^{n} \left( \alpha \frac{\|X\|^2}{n} - \zeta^2 \right) \frac{2\sigma_i^2}{(\sigma_i^2 + \alpha)^3}$$

Hence, a solution to $\left.\frac{\partial \langle \varepsilon^2 \rangle}{\partial \alpha}\right|_{\alpha = \alpha_{opt}} = 0$ is $\alpha_{opt} = \frac{n}{\|X\|^2} \zeta^2$ – irrespective of the singular value distribution.

The lowest achievable $\langle \varepsilon^2 \rangle$ is hence

$$\langle \varepsilon^2 \rangle|_{\alpha = \alpha_{opt}} = \left[ \frac{\|X\|^2}{n} \sum_{i=1}^{n} \frac{\alpha_{opt}^2}{(\sigma_i^2 + \alpha_{opt})^2} \right] + \left[ \zeta^2 \sum_{i=1}^{n} \frac{\sigma_i^2}{(\sigma_i^2 + \alpha_{opt})^2} \right]$$

$$= \sum_{i=1}^{n} \frac{\frac{\|X\|^2}{n} \alpha_{opt}^2 + \zeta^2 \sigma_i^2}{(\sigma_i^2 + \alpha_{opt})^2}$$

$$= \sum_{i=1}^{n} \frac{\frac{\|X\|^2}{n} \left( \frac{n}{\|X\|^2} \zeta^2 \right)^2 + \zeta^2 n \sigma_i^2}{\left( \sigma_i^2 + \frac{n}{\|X\|^2} \zeta^2 \right)^2}$$

$$= \frac{\|X\|^2}{n} \sum_{i=1}^{n} \frac{1}{1 + \frac{\sigma_i^2 \|X\|^2}{n\zeta^2}}$$

$$= \frac{\|X\|^2}{n} \sum_{i=1}^{n} \frac{1}{1 + \sigma_i^2 \rho}$$

Thus,

$$\chi = \frac{\langle \varepsilon^2 \rangle|_{\alpha=\alpha_{opt}}}{\|X\|^2} = \frac{1}{n} \sum_{i=1}^{n} \frac{1}{1 + \sigma_i^2 \rho} \quad (S3)$$

This is the result presented in Equation 2 of the main text. Note that for $R_{\text{eff}} = n$ this simplifies to the result derived in the previous section 4.1.

### 4.3 Random Channel Matrix

Let us now evaluate the general expression from Equation S3 for the special case of **H** being a random matrix. As discussed in the main text and in section 5.2, this does not directly correspond to the use of a random PCM coding sequence because of the presence of an unstirred field component. Nonetheless, this is an important benchmark.

First, we define the normalized singular values as $\sigma_i' = \frac{\sigma_i}{\sqrt{\frac{1}{n}\sum_{i=1}^{n} \sigma_i^2}}$, so $\sigma_i = \sigma_i' \sqrt{\frac{1}{n}\sum_{i=1}^{n} \sigma_i^2}$.

Thus, $\frac{\langle \varepsilon^2 \rangle|_{\alpha=\alpha_{opt}}}{\|X\|^2} = \frac{1}{n}\sum_{i=1}^{n} \frac{1}{1+\sigma_i^2 \rho} = \frac{1}{n}\sum_{i=1}^{n} \frac{1}{1+a\sigma_i'^2}$ with $a = \frac{\rho}{n}\sum_{i=1}^{n} \sigma_i^2$.

For a random matrix **H**, assuming finite-size effects are negligible, the probability distribution of $\sigma_i'$ is the well-known quarter circle law[3]: $p(\sigma_i') = \frac{1}{\pi}\sqrt{4 - \sigma_i'^2}$.

Thus,

$$\chi = \frac{\langle \varepsilon^2 \rangle|_{\alpha=\alpha_{opt}}}{\|X\|^2} = \int_0^2 P(\sigma_i') \frac{1}{1+a\sigma_i'^2} \, d\sigma_i'$$

$$= \frac{1}{\pi} \int_0^2 \frac{\sqrt{4-\sigma_i'^2}}{1+a\sigma_i'^2} \, d\sigma_i'$$

$$= \frac{\sqrt{4a+1} - 1}{2a}$$

Substituting $a = \frac{\rho}{n}\sum_i \sigma_i^2$ then yields

$$\chi = \frac{\sqrt{4\frac{\rho}{n}\sum_{i=1}^{n} \sigma_i^2 + 1} - 1}{2\frac{\rho}{n}\sum_{i=1}^{n} \sigma_i^2} \quad (S4)$$

In Figure S3 we confirm the validity of Equation S4 with numerically generated random matrices since, as stated previously, the use of a random coding sequence does not yield a channel matrix following Gaussian statistics.

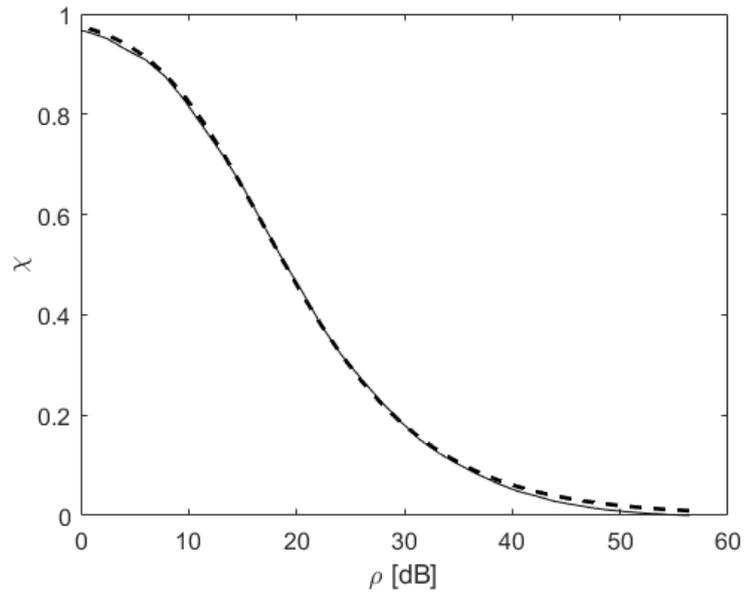

*Figure S3.* Comparison in simulation of the lowest achievable normalized mean squared error as a function of SNR. The continuous line is the reconstruction error on unseen data generated numerically for channel matrices with Gaussian statistics. The dashed line corresponds to the theoretical prediction from Equation S4.

## 5  Further Analysis of Experimental Channel Matrices

### 5.1  Visualization of Example Channel Matrices

Experimentally measured channel matrices corresponding to a random coding sequence and a coding sequence tailored to maximize the effective rank are displayed in Figure S4. Upon visual inspection, both matrices appear to have seemingly random entries. This observation confirms that the tailored coding sequence does not yield a trivial channel matrix that could have been predicted.

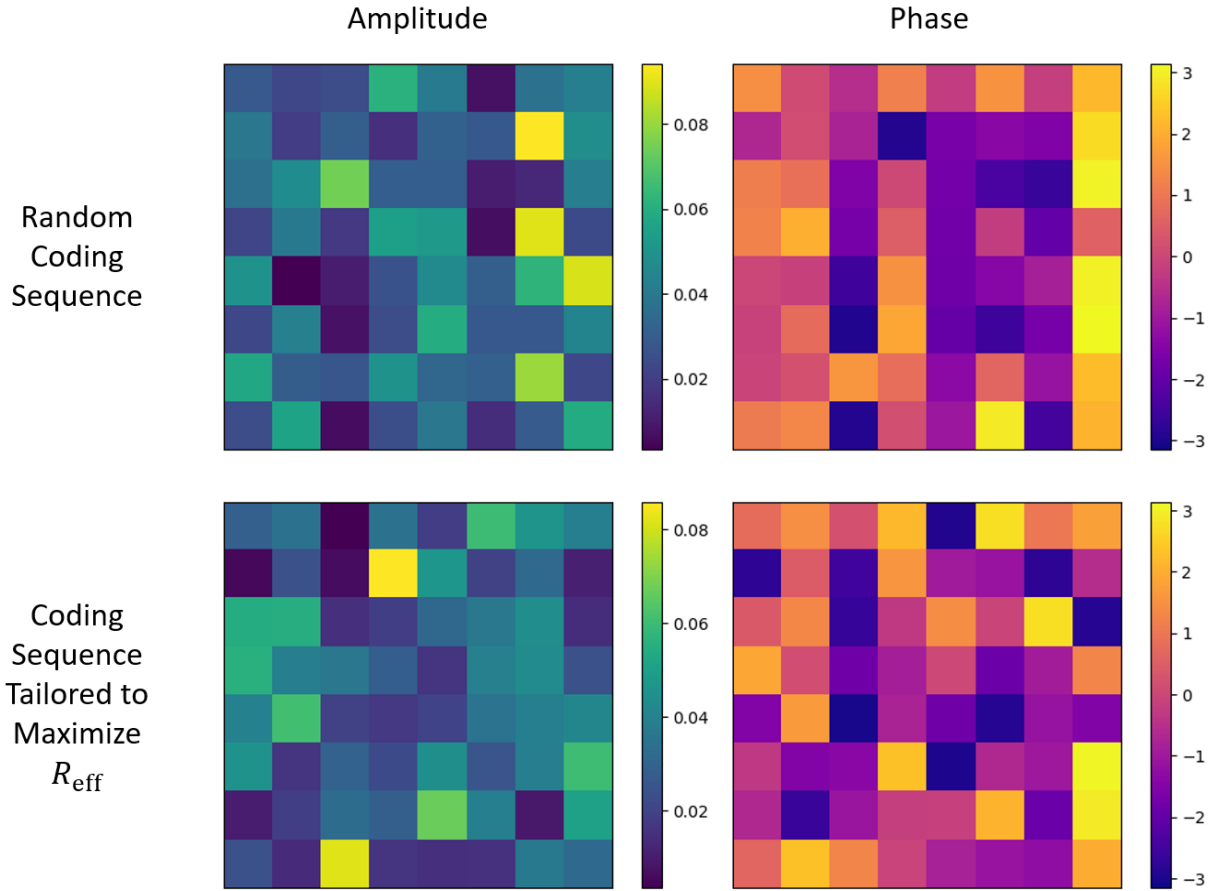

*Figure S4. Amplitude and phase of two channel matrices, one corresponding to a random coding sequence (top row) and one tailored to maximize the effective rank (bottom row).*

### 5.2  Distribution of Singular Values of Random-Coding-Sequence Channel Matrices

We evaluate the distribution of the normalized singular values $\sigma'_i = \frac{\sigma_i}{\sqrt{\frac{1}{n}\sum_{i=1}^{n}\sigma_i^2}}$ of 250 experimentally measured channel matrices corresponding to different random coding sequences.

First, we display in Figure S5a the singular value distribution obtained for 250 simulated random matrices. In contrast to the quarter circle law which is indicated for reference in red, the obtained distribution has eight distinct ripples. We attribute this to finite-size effects since the considered matrix dimension is $8 \times 8$. The effect is reminiscent of the "crystallization" of transmission eigenvalues observed in Figure 3a in Ref.[4]. Our use of the quarter circle law in Section 4.3 must thus be seen as an approximation; yet the observed agreement in Figure S3 is very good. We hypothesize that the deviations from the quarter circle law compensate each other.

Second, we display in Figure S5b the singular value distribution obtained for 250 experimentally measured channel matrices corresponding to distinct random PCM coding sequences. The most notable difference is the detachment of the strongest singular value from the rest. Upon careful inspection this is also observed in the average singular value distribution in Figure 2a of the main text. These deviations from the behavior of truly random matrices are due to the presence of an unstirred field component in the PCM. Consequently, the effective rank of the experimental random-coding-sequence channel matrices ($5.7 \pm 0.3$) is lower than for a truly random matrix ($6.5 \pm 0.2$), as reported in the main text.

Third, to confirm the above interpretation, we subtract the unstirred component from the channel matrix: $\mathbf{H} - \langle\mathbf{H}\rangle$, where the average is taken over all 250 realizations. The singular value distribution obtained only for the stirred components of the experimental channel matrices is displayed in Figures S5c and strongly resembles that of a truly random matrix in Figure S5a. The corresponding effective rank ($6.4 \pm 0.3$) is also comparable to that of a random matrix. Hence, the unstirred field component is indeed responsible for the deviations from the behavior of a truly random matrix.

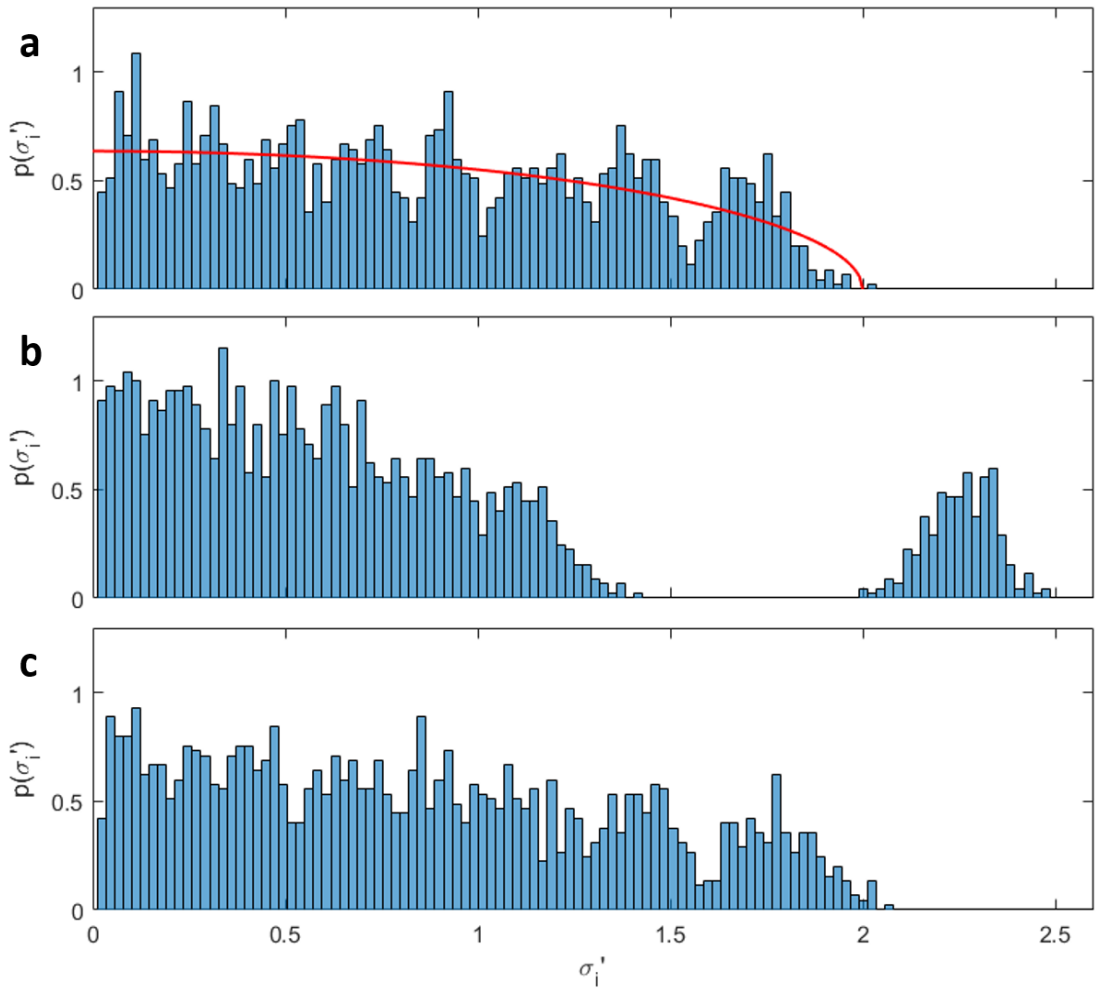

*Figure S5.* Distribution of singular values evaluated based on 250 realizations of $8 \times 8$ matrices. *a,* Numerically generated random Gaussian matrices. The red line indicates the quarter circle law for reference. *b,* Experimental channel matrices obtained with random coding sequences of the PCM. *c,* Stirred components of the experimental channel matrices obtained with a random PCM coding sequence: $\mathbf{H} - \langle\mathbf{H}\rangle$, where $\langle\mathbf{H}\rangle$ captures the unstirred components of $\mathbf{H}$.

Finally, we visualize for reference the presence of the unstirred component in Figure S6: the clouds of transmissions measured at the working frequency between each TX-RX pair for different random PCM configurations (gray dots) are not centered on the origin. We also indicate in color (see legend) the transmissions measured for the coding sequence tailored to maximize $R_{\mathrm{eff}}$.

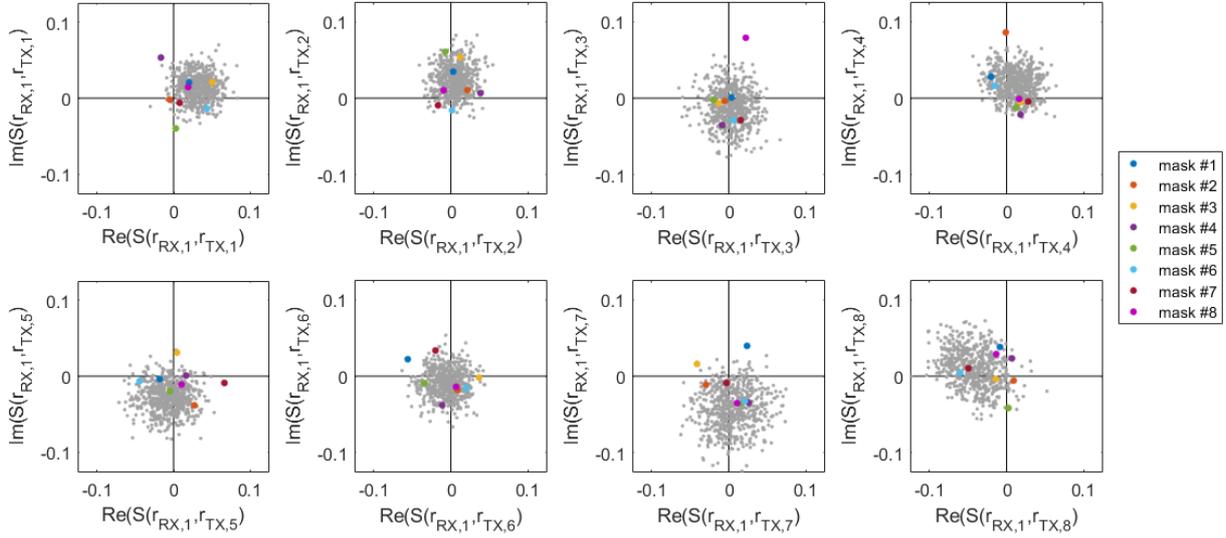

*Figure S6. Visualization of stirred and unstirred component of the field inside the PCM. For each transmission $S(\mathbf{r}_{\mathrm{RX},1}, \mathbf{r}_{\mathrm{TX},i})$ at the working frequency between the ith TX antenna at position $\mathbf{r}_{\mathrm{TX},i}$ and the RX antenna at position $\mathbf{r}_{\mathrm{RX},1}$, we plot in gray the complex values obtained with random PCM coding sequences. The fact that the clouds of gray points are not centered on the origin indicates that there is an unstirred field component. The transmissions measured for the coding sequence tailored to maximize the effective rank are color-coded (see legend).*

## 6 Effective Rank of Random Matrices

In this section, we evaluate the expected value of the effective rank of an uncorrelated Gaussian random matrix of size $n \times n$.

The starting point is the definition of the effective rank as given in Equation 3 in the main text:

$$R_{\mathrm{eff}} = e^{-\sum_{i=1}^{n} \frac{\sigma_i}{\sum_{i=1}^{n} \sigma_i} \ln\left(\frac{\sigma_i}{\sum_{i=1}^{n} \sigma_i}\right)}$$

Next, we substitute $\sigma_i = a\sigma'_i$, where $a = \sqrt{\frac{1}{n}\sum_{i=1}^{n} \sigma_i^2}$. Since $\frac{\sigma_i}{\sum_{i=1}^{n} \sigma_i} = \frac{a\sigma'_i}{\sum_{i=1}^{n} a\sigma'_i} = \frac{\sigma'_i}{\sum_{i=1}^{n} \sigma'_i}$, this yields

$$R_{\mathrm{eff}} = e^{-\sum_{i=1}^{n} \frac{\sigma'_i}{\sum_{i=1}^{n} \sigma'_i} \ln\left(\frac{\sigma'_i}{\sum_{i=1}^{n} \sigma'_i}\right)}$$

Using the fact that the distribution of $\sigma'_i$ is the quarter circle law $p(\sigma'_i) = \frac{1}{\pi}\sqrt{4 - \sigma'^2_i}$ for large $n$, we first evaluate $\langle \sum_{i=1}^{n} \sigma'_i \rangle$.

$$\langle \sum_{i=1}^{n} \sigma'_i \rangle = n \int_0^2 \sigma'_i p(\sigma'_i) \, d\sigma'_i = \frac{n}{\pi} \int_0^2 \sigma'_i \sqrt{4 - \sigma'^2_i} \, d\sigma'_i = \frac{8}{3\pi} n$$

Since we observe that the distribution of $\sum_{i=1}^{n} \sigma_i'$ is very narrow around a single peak (see Figure S7a), we make the following approximation:

$$\langle R_{\text{eff}} \rangle = \left\langle \exp\left(-\sum_{i=1}^{n} \frac{\sigma_i'}{\frac{8}{3\pi}n} \ln\left(\frac{\sigma_i'}{\frac{8}{3\pi}n}\right)\right)\right\rangle = \left\langle \exp\left(-\sum_{i=1}^{n} \frac{3\pi}{8n}\sigma_i' \ln\left(\frac{3\pi}{8n}\sigma_i'\right)\right)\right\rangle$$

Now, let us first evaluate the expected value of the exponent.

$$\langle \ln(R_{\text{eff}}) \rangle = \left\langle -\sum_{i=1}^{n} \frac{\sigma_i'}{\frac{3\pi}{8n}\sigma_i'} \ln\left(\frac{3\pi}{8n}\sigma_i'\right) \right\rangle$$

$$= -n \int_0^2 \frac{3\pi}{8n}\sigma_i' \ln\left(\frac{3\pi}{8n}\sigma_i'\right) p(\sigma_i') \, d\sigma_i'$$

$$= -\frac{3}{8} \int_0^2 \sigma_i' \ln\left(\frac{3\pi}{8n}\sigma_i'\right) \sqrt{4 - \sigma_i'^2} \, d\sigma_i'$$

$$= -\frac{3}{8} \int_0^2 \sigma_i' \left[\ln\left(\frac{3\pi}{8n}\right) + \ln(\sigma_i')\right] \sqrt{4 - \sigma_i'^2} \, d\sigma_i'$$

$$= -\frac{3}{8}\left[\left(\ln\left(\frac{3\pi}{8n}\right) \int_0^2 \sigma_i' \sqrt{4 - \sigma_i'^2} \, d\sigma_i'\right) + \left(\int_0^2 \sigma_i' \ln(\sigma_i') \sqrt{4 - \sigma_i'^2} \, d\sigma_i'\right)\right]$$

$$= -\frac{3}{8}\left[\left(\ln\left(\frac{3\pi}{8n}\right)\frac{8}{3}\right) + \left(\frac{8}{9}(\ln(64) - 4)\right)\right]$$

$$= -\ln\left(\frac{3\pi}{8n}\right) - \frac{1}{3}(\ln(64) - 4)$$

Since we observe in Figure S7b that the distribution of $\ln(R_{\text{eff}})$ is also very narrow around a single peak, we approximate $\langle R_{\text{eff}} \rangle \approx e^{\langle \ln(R_{\text{eff}}) \rangle}$. Simplifying $\langle R_{\text{eff}} \rangle = e^{-\ln\left(\frac{3\pi}{8n}\right) - \frac{1}{3}(\ln(64) - 4)}$ we find

$$\boxed{\langle R_{\text{eff}} \rangle = \frac{8 e^{\frac{1}{3}(4 - \ln(64))}}{3\pi} n \approx 0.805 n} \quad (S5)$$

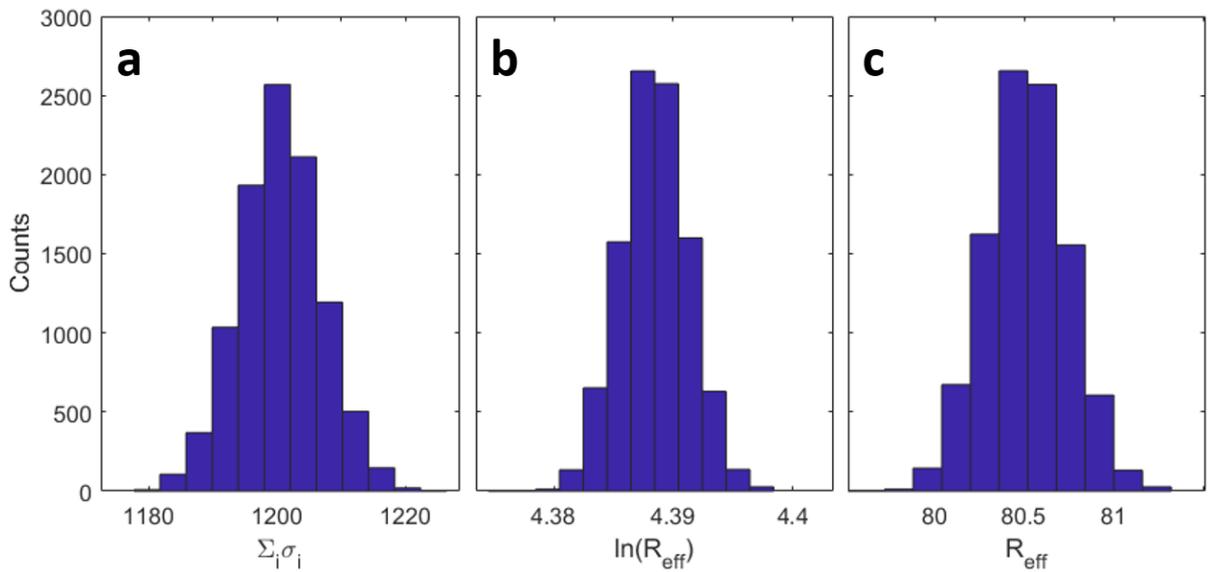

**Figure S7.** *Histograms of three quantities obtained for 10000 realizations of numerically generated $100 \times 100$ random matrices: **a,** sum of the singular values; **b,** natural logarithm of the effective rank; **c,** effective rank.*

The validity of Equation S5 is confirmed in a numerical study presented in Figure S8 below. Surprisingly, Equation S5 also explains the observed behavior for very small $n$ – even though the derivation of Equation S5 is based on the quarter circle law which applies only to cases with large $n$.

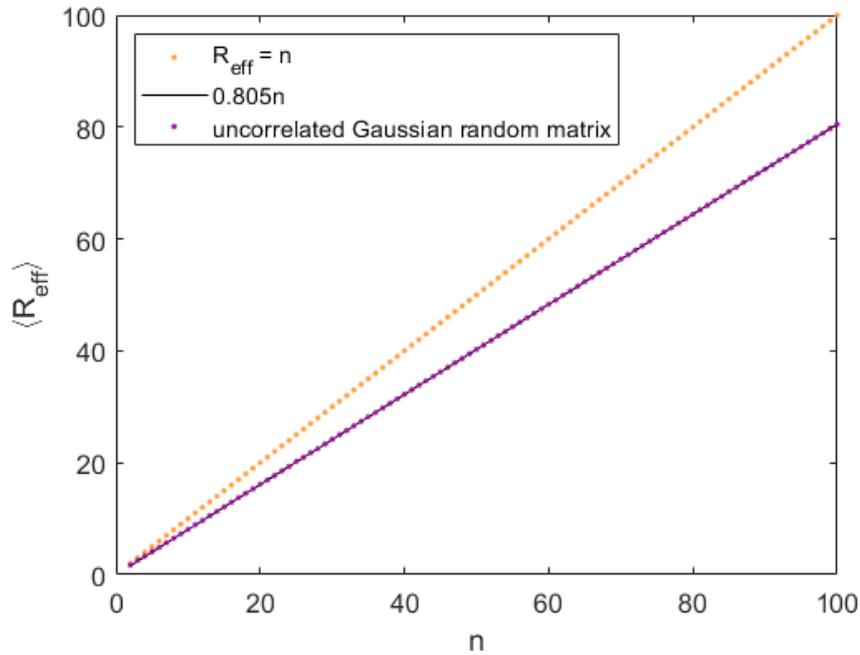

*Figure S8. Dependence of the average effective rank of an uncorrelated Gaussian $n \times n$ random matrix on $n$. For reference, the curves $R_{\text{eff}} = n$ and $R_{\text{eff}} = 0.805n$ are shown, the latter corresponding to Equation S5.*

**References**

[1] A. N. Tikhonov, *Soviet Math. Dokl.* **1963**, *4*, 1035.
[2] M. Gockenbach, *Linear Inverse Problems and Tikhonov Regularization*, The Mathematical Association Of America, **2016**.
[3] V. Marćenko, L. Pastur, *Sbornik Math.* **1967**, *1*, 457.
[4] Z. Shi, A. Z. Genack, *Phys. Rev. Lett.* **2012**, *108*, 043901.